\documentclass[twocolumn,aps,floatfix]{revtex4}
\usepackage{amssymb}
\usepackage{graphicx}
\usepackage{pstricks}

\begin{document}

\title{Constructing classical field for a Bose-Einstein condensate in arbitrary
       trapping potential; quadrupole oscillations at nonzero temperatures}

\author{Tomasz Karpiuk,$^1$ Miros{\l}aw Brewczyk,$^2$ Mariusz Gajda,$^{3,5}$ and 
        Kazimierz Rz\c a\.zewski$\,^{4,5}$}                          

\affiliation{\mbox{$^1$ Centre for Quantum Technologies, National University 
              of Singapore,} \\ 
\mbox{3 Science Drive 2, Singapore 117543, Singapore} \\
\mbox{$^2$ Wydzia{\l} Fizyki, Uniwersytet w Bia{\l}ymstoku, 
            ulica Lipowa 41, 15-424 Bia{\l}ystok, Poland}  \\
\mbox{$^3$ Instytut Fizyki PAN, Aleja Lotnik\'ow 32/46, 02-668 Warsaw, Poland} \\
\mbox{$^4$ Centrum Fizyki Teoretycznej PAN, Aleja Lotnik\'ow 32/46, 02-668 Warsaw, 
           Poland}  \\
\mbox{$^5$ WMP-SN\'S, UKSW, Warsaw, Poland}    }             

\date{\today}

\begin{abstract}
We optimize the classical field approximation of the version described in 
J. Phys. B {\bf 40}, R1 (2007) for the oscillations of a Bose gas trapped in a
harmonic potential at nonzero temperatures, as experimentally investigated by Jin {\it et al.} 
[Phys. Rev. Lett. {\bf 78}, 764 (1997)]. Similarly to experiment, the system response to 
external perturbations strongly depends on the initial temperature and on the symmetry of 
perturbation. While for lower temperatures the thermal cloud follows the condensed part, 
for higher temperatures the thermal atoms oscillate rather with their natural frequency,
whereas the condensate exhibits a frequency shift toward the thermal cloud frequency 
($m=0$ mode), or in the opposite direction ($m=2$ mode). In the latter case, for
temperatures approaching critical, we find that the condensate begins to oscillate
with the frequency of the thermal atoms, as in the $m=0$ mode.
A broad range of frequencies of the perturbing potential is considered.

\end{abstract}

\maketitle

\section{Introduction}
Experiments with atomic Bose-Einstein condensates driven by an external perturbation
have allowed verification of the mean-field description of the condensed phase in a dynamical
regime where the system responds by collective motion.
Typically, a periodic perturbation (with a particular symmetry) of a trapping potential
was used to excite the Bose gas \cite{Cornell-zero,Ketterle-zero,Cornell-osc,Ketterle-osc},
although other kinds of trap distortion leading, for instance, to scissors mode excitations
\cite{Foot} or the transverse monopole mode excitation in an elongated condensate\cite{Dalibard}
were also tried. The investigation of the low energy collective modes of the condensate in 
the zero-temperature limit \cite{Cornell-zero,Ketterle-zero} has revealed that the mean-field 
description of that system (i.e., based on the Gross-Pitaevskii equation) works well 
\cite{theory-zero}. However, when the study of low-lying excitations was extended to include 
the measurement of frequencies and damping rates as a function of temperature 
\cite{Cornell-osc,Ketterle-osc}, it became clear that a new theory of an interacting Bose gas 
at nonzero temperatures is required.

The JILA experiment of Ref. \cite{Cornell-osc} showed two effects. First, it was found that
two collective modes with different symmetries (quadrupole modes with angular momenta
equal to $m=0$ and $m=2$) behave in qualitatively different ways. When the temperature 
increases, they exhibit a frequency shift in opposite directions. Moreover, for the $m=0$ mode
a rather sudden upward shift is observed, suggesting the existence of a characteristic
temperature which is approximately $0.65$ of the critical temperature for the corresponding
ideal gas. Secondly, the damping of the collective oscillations turned out to be dramatically
dependent on temperature, showing that the condensate modes are damped even faster than the
noncondensed fraction while approaching the critical temperature.  All these puzzling 
findings triggered a lot of theoretical work and after a few years resulted
in the development of Zaremba-Nikuni-Griffin formalism \cite{ZNG,Zaremba-osc} and the
second-order quantum field theory \cite{Burnett-osc,Morgan} for a Bose gas.  Recently, 
another attempt to describe finite-temperature properties of low-lying collective modes 
was undertaken in Ref. \cite{Blakie-osc} within the approximation called the projected 
Gross-Pitaevskii equation.

The Zaremba-Nikuni-Griffin formalism \cite{Zaremba-osc} applied to the results of the JILA 
experiment gives relatively good agreement. Also the calculations based on the
second-order quantum field theory \cite{Burnett-osc} show a good agreement with experimental
data. However, already these two approaches differ when considering their fundamentals. 
For example, the first one neglects the phonon character of the low-lying energy modes,
the anomalous average, as well as the Beliaev processes. Regarding the JILA experiment, for the 
$m=0$ mode, the Zaremba-Nikuni-Griffin method predicts  an additional branch of condensate
frequencies, so far not observed experimentally. No such branch is found within the second-order
theory of Ref. \cite{Burnett-osc}. Furthermore, this approach predicts a single frequency
of condensate response for a particular temperature, regardless of driving frequency, and hence the 
notion of a natural condensate
frequency is meaningful. This strongly differs from  what is reported in Jackson and Zaremba 
\cite{Zaremba-osc}, where the condensate response depends on the driving frequency of the whole
system (condensed and noncondensed parts). A comprehensive discussion of 
both approaches in the context of JILA experiment can be found in a recent review article
\cite{Proukakis}.  
Another formalism, based on the
projected Gross-Pitaevskii equation, used to model the JILA experiment \cite{Cornell-osc}
produces good agreement with experimental data up to $0.65 T_c$, and for the $m=2$ mode at
higher temperatures. However, it fails to predict the sudden upward frequency shift for
the $m=0$ mode at this temperature \cite{Blakie-osc}.
This failure is perhaps related to the way the cutoff parameter (which splits the space
of modes into the highly occupied ones that are described in terms of the classical field, and 
the others that are sparsely occupied and require, in principle, a quantum treatment) is chosen. 
The details of the splitting procedure within the projected Gross-Pitaevskii equation
approach are discussed in Ref. \cite{Advances}.

In this paper we apply the classical field approximation in the version described in
Ref. \cite{przeglad} to the case of a trapped interacting Bose gas at nonzero
temperatures driven by an external perturbation as in the experiment of Ref. 
\cite{Cornell-osc}. The main purpose of this work is to check whether the
classical field approximation is able to reproduce, at least qualitatively, the
findings of JILA experiment. This is an especially important task because of the recently reported
failure \cite{Blakie-osc} to explain the behavior of the $m=0$ mode 
within the projected Gross-Pitaevskii equation method, which is conceptually very
close to our classical field approximation. On the other hand, although the other 
existing theories \cite{Zaremba-osc} and \cite{Burnett-osc} both lead to relatively good
agreement with the experiment \cite{Cornell-osc}, they somehow contradict each other conceptually
as discussed in the previous paragraph. It would be nice to have an alternative
view of the processes going on in the perturbed Bose gas. Finally, the approaches
\cite{Zaremba-osc} and \cite{Burnett-osc} have some difficulties in describing the
dynamics of the thermal cloud, especially in the $m=2$ mode. In fact, no predictions for
thermal component frequencies in the $m=2$ mode are given in \cite{Zaremba-osc} or 
\cite{Burnett-osc}. Simultaneously, within the projected Gross-Pitaevskii equation
method, at higher temperatures, the thermal cloud (in fact, for both the $m=0$ and $m=2$ modes) oscillates 
at frequencies much lower than in experiment.

The classical field approximation has already been applied in static and dynamical
regimes for a uniform and harmonically trapped systems (for a review, see 
\cite{przeglad}). This approach was used to investigate the thermodynamics of
an interacting gas \cite{rapid,JPB} as well as dynamical processes like
 the photoassociation of molecules \cite{photo}, the dissipative dynamics
of a vortex \cite{vortex1}, the superfluidity in ring-shaped traps \cite{super},
or the thermalization in spinor condensates \cite{spinor}. 
The classical field approximation was also tested at a quantitative level when e.g. 
the Bogoliubov-Popov quasiparticle energy spectrum in a uniform Bose 
gas was obtained \cite{JPB} or when the process of splitting of doubly quantized 
vortices in dilute Bose-Einstein condensates \cite{vortex2} was studied.

The paper is organized as follows: 
In Sec. \ref{method} we describe the classical field approximation for a trapped Bose
gas with particular attention on how the equilibrium state is obtained and on
the quality of the solution in comparison with the equilibrium states calculated within 
the self-consistent Hartree-Fock method.
Sec. \ref{mode0} discusses the results (response frequencies and damping rates) for the 
quadrupole $m=0$ collective mode for parameters as in JILA experiment \cite{Cornell-osc} 
while in Sec. \ref{mode2} we do the same for $m=2$ excitation.
Finally, we conclude in Sec. \ref{concl}.

\section{Classical field approximation for a trapped Bose gas}
\label{method}

\subsection{Formalism}
\label{formalism}

A good starting point to introduce the classical field approximation is the usual Heisenberg 
equation of motion for the bosonic field operator $\hat {\Psi }({\bf r},t)$ which annihilates 
an atom at point ${\bf r}$ and time $t$. The field operator $\hat {\Psi }({\bf r},t)$ fulfills 
standard commutation relations:
\begin{equation}
\left[ {\hat {\Psi }({\rm {\bf r}},t),\hat {\Psi }^+({\rm {\bf r}'},t)}
\right]=\delta ({\rm {\bf r}}-{\rm {\bf r}'})
\label{comrel}
\end{equation}
with other equal time commutation relations for $[\hat {\Psi},\hat {\Psi}]$ and
$[\hat {\Psi}^+,\hat {\Psi}^+]$ being zeros. The equation of motion reads:
\begin{eqnarray}
&&i\hbar \frac{\partial}{\partial t} \hat {\Psi }({\rm {\bf r}},t) =
\left[ -\frac{\hbar^2}{2m} \nabla^2 + V_{tr}({\rm {\bf r}},t)    \right]
\hat {\Psi }({\rm {\bf r}},t)   \nonumber  \\
&&+ g\, \hat{\Psi }^+({\rm {\bf r}},t) \hat {\Psi }({\rm {\bf r}},t)
\hat {\Psi }({\rm {\bf r}},t)   \,.
\label{Heisenberg}
\end{eqnarray}
Here, we assume the time-dependent trapping potential $V_{tr}$ and the usual contact interaction 
for colliding atoms. The coupling constant $g=4\pi \hbar^2 a /m$ is expressed in terms of 
the s-wave scattering length $a$.

Next, we expand the field operator $\hat {\Psi }({\bf r},t)$ in the basis of one-particle
wave functions $\psi_k({\bf r})$, where $k$ is a set of one-particle quantum numbers:
\begin{equation}
\hat {\Psi }({\rm {\bf r}},t) = \sum_k \psi_k({\bf r})  \hat {a}_k(t)
\label{expansion}
\end{equation}
Now, we assume that some of the modes used in the expansion (\ref{expansion}) are macroscopically
occupied and extend the original Bogoliubov idea \cite{Bogoliubov} by replacing all operators 
$\hat {a}_k(t)$ corresponding to these modes by c-numbers. When only macroscopically occupied modes
are considered, the field operator $\hat {\Psi }({\bf r},t)$ is turned into the complex wave function 
$\Psi ({\bf r},t)$ and the expansion (\ref{expansion}) takes the form:
\begin{equation}
\Psi ({\rm {\bf r}},t) = \sum_{k=0}^{k_{max}} \psi_k({\bf r}) a_k(t)  \,.
\label{expansion1}
\end{equation}
The upper index in the summation tells us that, indeed, the wave function $\Psi ({\bf r},t)$ 
is expanded only over a finite number of states, i.e. those which are macroscopically occupied. 
We call the wave function $\Psi ({\bf r},t)$ the classical field. This is analogous to the way 
that intense electromagnetic waves can be treated. In spite of consisting of photons, an intense light beam is well characterized
by the classical electric and magnetic fields. Since the experiments with dilute atomic gases are 
performed with millions of atoms it seems to be a plausible approximation to use the classical 
field to describe atoms in analogy with electric and magnetic fields for photons.

Obviously, the classical field obeys the following equation:
\begin{eqnarray}
&&i\hbar \frac{\partial}{\partial t} {\Psi }({\rm {\bf r}},t) =
\left[ -\frac{\hbar^2}{2m} \nabla^2 + V_{tr}({\rm {\bf r}},t)    \right]
{\Psi }({\rm {\bf r}},t)   \nonumber  \\
&&+ g\, {\Psi }^*({\rm {\bf r}},t) {\Psi }({\rm {\bf r}},t)
{\Psi }({\rm {\bf r}},t)   \,.
\label{CFequation}
\end{eqnarray}
We usually implement the cutoff parameter $k_{max}$ by solving the Eq. (\ref{CFequation})
on a rectangular grid using the Fast Fourier Transform technique. The spatial grid step
determines the maximal momentum per particle (and hence the energy) in the system whereas the
use of the Fourier transform implies a projection in momentum space.

The equation (\ref{CFequation}) looks like the usual Gross-Pitaevskii equation
describing the Bose-Einstein condensate at zero temperature. However, here the interpretation
of the complex wave function $\Psi ({\bf r},t)$ is different. It describes all the atoms
in the system, both those in a condensate and in a thermal cloud. Therefore, the question 
appears how to split the classical field $\Psi ({\bf r},t)$ into the condensed and noncondensed
fractions. For that we use the definition of a Bose-Einstein condensation proposed originally
by Penrose and Onsager \cite{POdef}. According to this definition the condensate is assigned
to the eigenvector corresponding to the dominant eigenvalue of a one-particle density matrix.
So, we built the one-particle density matrix in the following way:
\begin{equation}
\rho^{(1)}({\bf r},{\bf r}^{\,\prime};t) = \frac{1}{N}\, \Psi({\bf r},t)\, 
\Psi^*({\bf r}^{\,\prime},t)   \,,
\label{denmat}
\end{equation}
where $N$ is the total number of particles. However, and here comes the surprise -- since (\ref{denmat}) 
is just the spectral decomposition of a one-particle density matrix, this would imply that the classical 
field $\Psi ({\bf r},t)$ is the condensate wave function and that all atoms are in the condensate. 
To split the classical field into the condensed and thermal fractions one first realizes that the 
high energy solutions of Eq. (\ref{CFequation}) oscillate rapidly in time and space. On the 
other hand, the detection process is always performed with limited spatial and temporal resolution.
Therefore, it becomes clear that the measurement process with its limited resolution involves a kind of 
 averaging (coarse-graining) of Eq. (\ref{denmat}). Again, an analogy 
with electromagnetic waves is in place. A partially coherent light beam exhibits complicated 
spatial and temporal behavior on short scales, in fact too complicated to be typically measured. What is 
important are then
the correlation functions at long enough spatial and temporal separation, averaged over the smaller time 
intervals and space segments. Similarly here, the 
averaging over time and/or space of the one-particle density matrix (\ref{denmat}) results 
in a partial loss of the information contained in the classical field \cite{rapid}. In other words, 
the mixed state emerges out of the pure one.

In a typical experiment, what is measured is the column density along some direction. Hence, we also implement
the coarse-graining procedure
in our numerics in this manner as:
\begin{equation}
\bar{\rho}(x,y,x',y';t) = \frac{1}{N}  \int dz \, \Psi(x,y,z,t) \, \Psi^*(x',y',z,t)  \,.
\label{rhoave}
\end{equation}

Solving the eigenvalue problem for a coarse-grained density matrix (\ref{rhoave}) results in
a decomposition:
\begin{equation}
\bar{\rho}=\sum_k n_k \, \varphi_k(x,y,t) \, \varphi^*_k(x^{\,\prime},y^{\,\prime},t)  \,,
\end{equation}
where $n_k = N_k /N$ are the relative occupations of macroscopically occupied modes $\varphi_k$. 
Defining the dominant eigenvalue as $n_0$, the condensate wave function (normalized to
$N_0/N$) can be written as:
\begin{equation}
\psi_0(x,y,t) =  \sqrt{\frac{N_0}{N}}\, \varphi_0(x,y,t)   \,.
\end{equation}
All the other modes contribute to the thermal density which is, therefore, given by:
\begin{equation}
\rho_T(x,y,t) = \bar{\rho}(x,y,x,y;t) - |\psi_0(x,y,t)|^2  \,.
\end{equation}

The appropriateness of the averaging (\ref{rhoave}) for obtaining the large eigenvalues of a coarse-grained density matrix 
was verified by us in various ways. For example, we checked that for the 
classical field at equilibrium this procedure gives the same results 
as averaging over a long enough time. Prescription (\ref{rhoave}) also works at zero temperature 
(when all atoms are expected to be in a condensate), having been successfully tested in this respect in the demanding case of a lattice 
of bent vortices (see Ref. \cite{mcv} for further details).

According to the description given above the splitting of the system into condensed and 
noncondensed components is a result of Bose statistics, interaction, and the measurement 
process. Unlike the alternative approaches \cite{ZNG,Burnett-osc} we do not impose a
two-component character of the system  from the beginning. Also, our version of the classical field 
method is well suited to describe single realizations of the experiment since it corresponds to a 
microcanonical ensemble, as opposed to the competing approach \cite{CastinCFA}, which deals with canonical 
ensembles.

\subsection{Obtaining equilibrium states}
\label{eqstates}

\begin{figure}[htb]
\resizebox{3.2in}{2.1in} {\includegraphics{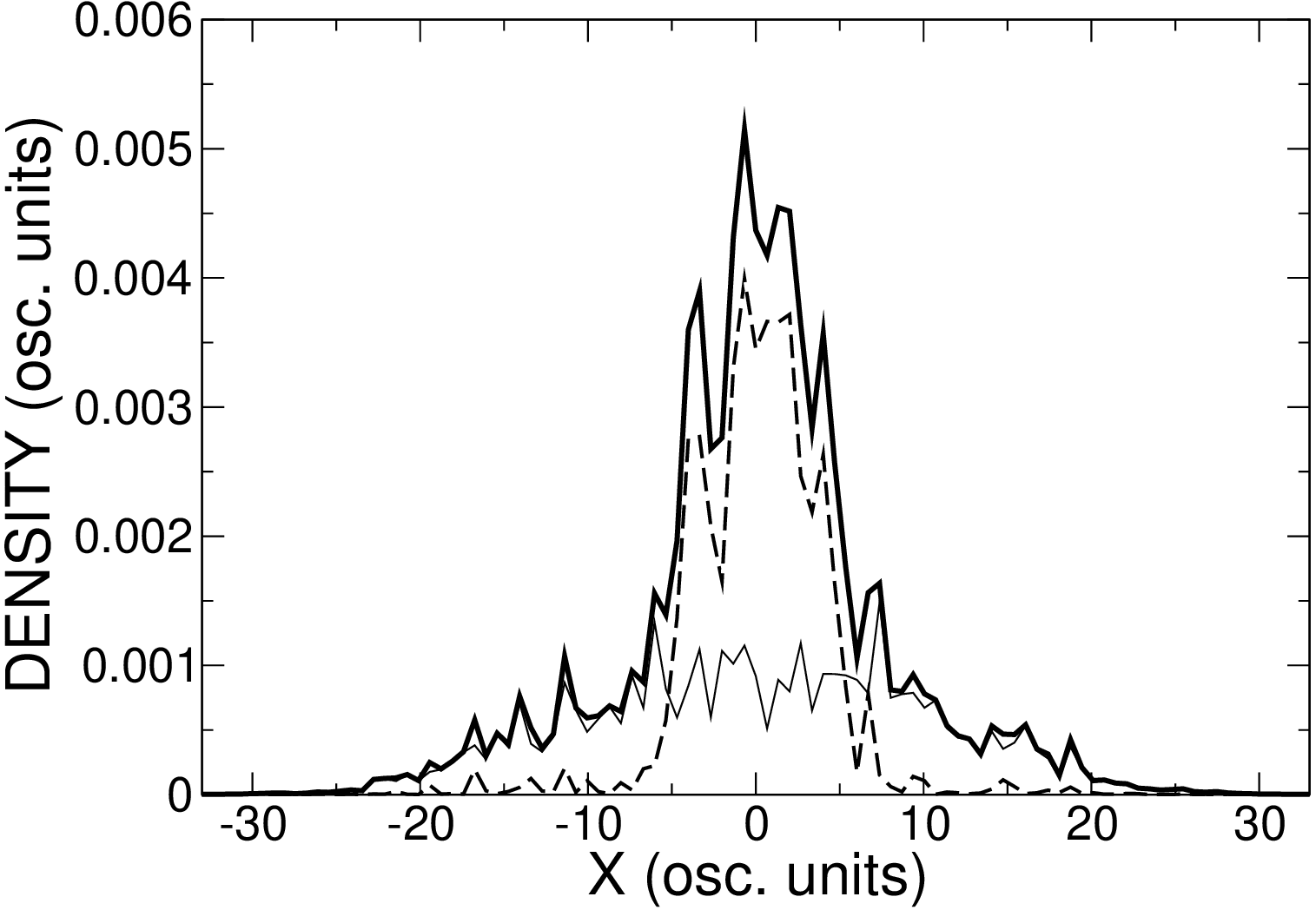}}
\caption{Cuts of the total (thick solid line), condensate (dashed line), and thermal (thin solid 
line) densities as obtained by integrating a one-particle density matrix along the direction of 
imaging and then applying the splitting procedure described in Sec. \ref{formalism}. Note the bimodal
character of the density distribution (here, the condensate fraction equals $0.3$).
The oscillatory unit of length is defined via the axial trap frequency: 
$\sqrt{\frac{\hbar}{m \omega_z}}$ and equals $0.565\, \mu m$.  }
\label{cfeq}
\end{figure}

An initial classical field is generated from the ground state solution of Eq.
(\ref{CFequation}) by adding appropriate random disturbance. An inspection of that equation 
indicates that only the product of the coupling constant $g$ and the total number of atoms $N$ enters it. 
Therefore, we normalize
the initial classical field $\Psi ({\bf r},t=0)$ to unity. The norm of $\Psi ({\bf r},t)$
is then one of the constants of motion of Eq. (\ref{CFequation}). Another constant of motion is 
the total energy.  Such an initial state is then propagated according to Eq. (\ref{CFequation}) 
until the constituent energies (kinetic, trap, and interaction) cease to change systematically in time 
and exhibit only fluctuations. In this way the classical field at thermal equilibrium,
corresponding to the particular values of $gN$ and $E_{tot}/N$, is obtained. An example is
given in Fig. \ref{cfeq}, where we plot cuts of the total density, the condensate, and the thermal 
densities according to the prescription detailed in the previous section. Here, the equilibrium 
classical field describes a degenerate $^{87}$Rb Bose gas in a magnetic trap with frequencies 
$\omega_{\bot} \equiv \omega_{x,y} = 2\pi \times 129\,$Hz and $\omega_z = 2\pi \times 365\,$Hz
as in the experiment of Ref. \cite{Cornell-osc}. Other parameters are: $gN=2911.9$ and
$E_{tot}/N=21.2$ in units of $\hbar \omega_z (\hbar / m \omega_z)^{3/2}$ and $\hbar \omega_z$,
respectively. They result in a condensate fraction $n_0=0.3$. A bimodal distribution
(thick solid line) is clearly visible.

\begin{figure}[htb]
\resizebox{3.2in}{2.1in} {\includegraphics{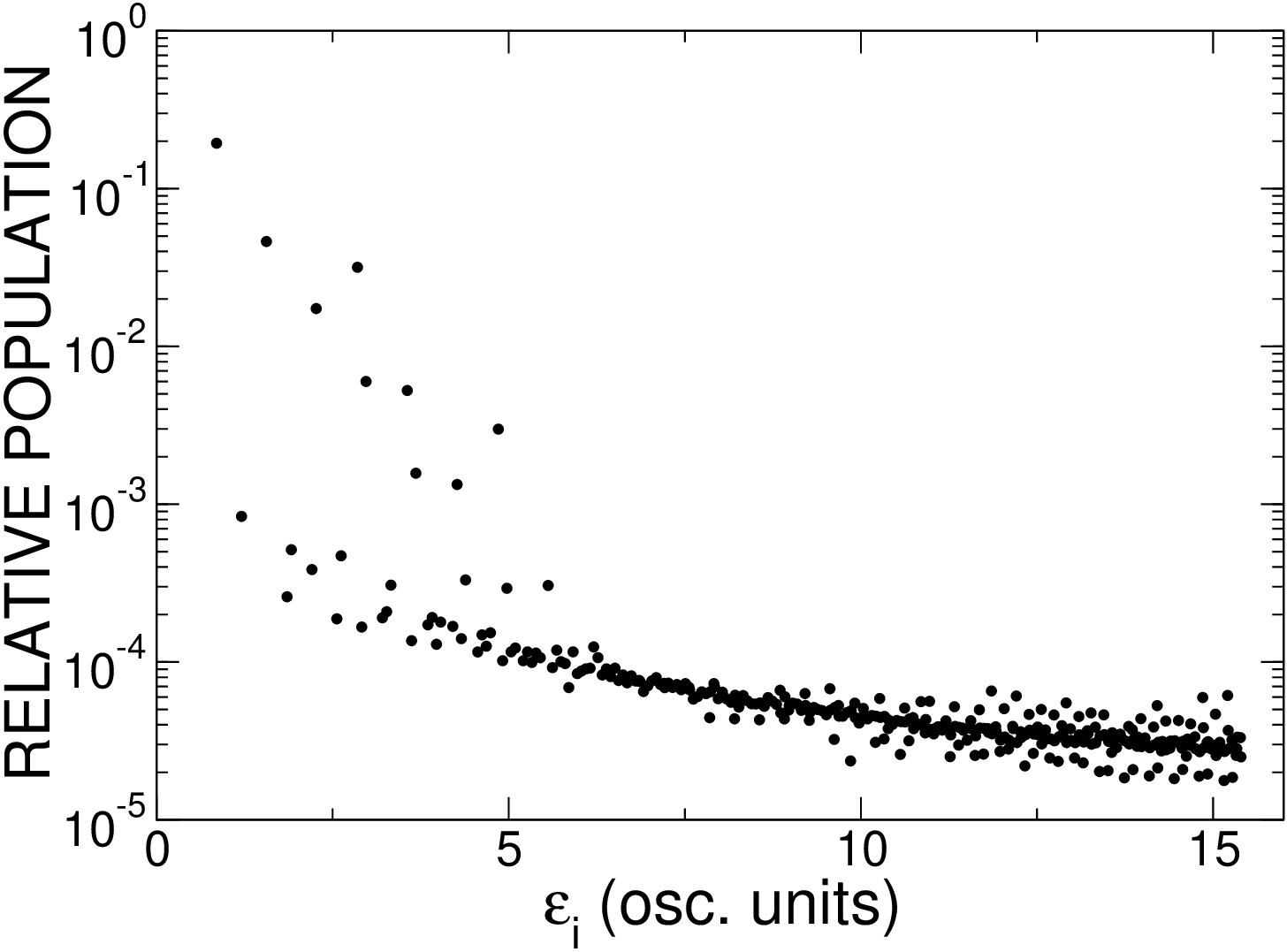}}
\caption{Relative populations of various harmonic oscillator states.
The energy cutoff equals $p_{max}^2/m$, where $p_{max}$ is the momentum
cutoff. }
\label{relpop}
\end{figure}

\begin{figure}[htb]
\resizebox{3.2in}{2.1in} {\includegraphics{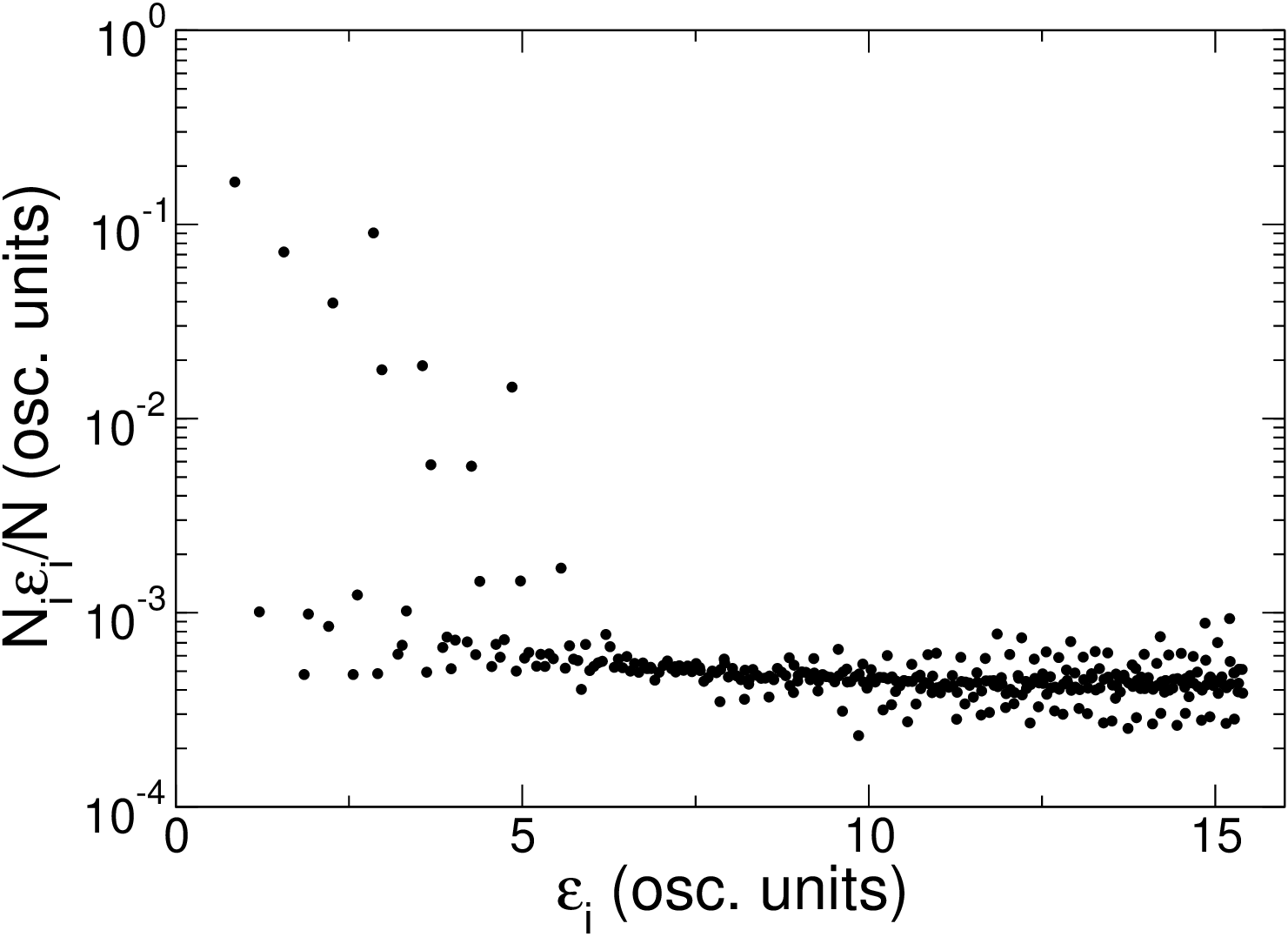}}
\caption{Relative populations multiplied by the state energies for various harmonic 
oscillator states. The figure clearly shows that for higher energy states the equipartition 
of an energy is established. Therefore, the higher energy states become quasiparticle modes.
The fraction $k_B T/N$ equals approximately $4.62 \times 10^{-4}$. }
\label{kTN}
\end{figure}

Now we have to solve the problem how to find the number of particles assigned to the
classical field at equilibrium and how to determine the temperature of the system.
This can be done in two ways. The first is given here, the second in Sec.~\ref{HFmethod}.
To begin with, having the classical field at equilibrium one can
project the field $\Psi ({\bf r},t)$ on the harmonic oscillator states obtaining in this way
the relative populations of these states. Fig. \ref{relpop} shows relative populations for
$gN=1811.9$ and $E_{tot}/N=10.0$ as a function of the harmonic oscillator states' energy
(precisely, the time average over $274\,$ms is plotted). The maximal one-particle
mode energy is determined by the momentum cutoff $p_{max}$ as $p_{max}^2/m$.
An important observation is made when one looks at the energy accumulated in harmonic 
oscillator states, see Fig. \ref{kTN}.  
For higher energy states the product $n_i \, \varepsilon_i$ (where $n_i$ and 
$\varepsilon_i$ are the relative population and the harmonic oscillator state energy, 
respectively) becomes constant. On the other hand, for highly occupied modes (i.e., modes 
satisfying $\epsilon_i - \mu \lesssim k_B T$, where $\mu$ is the chemical potential) the quantum 
Bose-Einstein distribution reduces to the classical equipartition distribution. For the classical field
studied here the equipartition extends all the way to the cutoff energy:
\begin{equation}
N_i \, (\epsilon_i - \mu) = k_B T \,,
\end{equation}
which can be written equivalently as
\begin{equation}
n_i \, (\epsilon_i - \mu) = k_B T / N   \,.
\label{qppop}
\end{equation}
Therefore, since energy equipartition is established, the higher energy harmonic oscillator states become the 
quasiparticle modes. In Fig. \ref{kTN} we determine the ratio $k_B T/N$ to be
 $4.62 \times 10^{-4}$. Having the ratio $k_B T/N$ we now find from Eq. (\ref{qppop})
the relative populations of high energy quasiparticle modes. In particular, we get the relative
population of the least occupied modes which belong to the classical field. In the example here,
they are $2.99 \times 10^{-5}$ (based on formula (\ref{qppop}) with $k_B T/N$ obtained above). In Ref. \cite{EW}, arguments are given that the best match between this method and the 
ideal Bose gas occurs when the occupation of the least occupied mode is $N_{cut}=0.46$. 
These are based on the comparison between the probability distribution of the ideal Bose gas and its classical field counterpart.
Assuming, then that the average number of atoms in this least-occupied mode is 
$0.46$ we can retrieve the total number of atoms in the system and its temperature separately. 
For the example here, these are $N=15342$ and $T=124.1\,$nK. 
Since no data on the population of the cutoff mode are available for a weakly interacting 
Bose gas considered here we compare in Table \ref{tab0} the temperature of the system and the 
total number of atoms for various values of $N_{cut}$. 

\begin{table}[htb]
 \begin{tabular}{|c|c|c|}
\hline
  $N_{cut}$ & $T$[nK] & $N$ \\
\hline \hline
0.44 & 119.5 & 14778    \\
0.46 & 124.1 & 15342   \\
0.48 & 130.4 & 16121   \\
0.50 & 135.8 & 16793   \\
0.52 & 141.2 & 17465   \\
\hline
 \end{tabular} 
\caption{Temperature and the total number of atoms obtained by projecting the classical field 
on the harmonic oscillator states of the example described in Sec. \ref{eqstates} for different values of cutoff parameter.}
\label{tab0}
\end{table}

\subsection{Comparison with the self-consistent Hartree-Fock model}
\label{HFmethod}

Another approach to obtain the number of atoms and the temperature of the system
described by the classical field at equilibrium is based on the self-consistent Hartree-Fock
method \cite{Pethick}. Since this method works well for a Bose
gas at equilibrium as verified experimentally in \cite{Aspect}, a comparison between
this model and the classical field approximation should be instructive. The Hartree-Fock
description is defined by the following set of equations \cite{Pethick}:
\begin{eqnarray}
&&n_c({\bf r}) = \frac{1}{g}\left[ \mu - V_{tr}({\bf r}) - 2\, g\, n_{th}({\bf r}) \right]
\label{SCHF1}  \\
&&f({\bf r},{\bf p}) = \left( e^{ [{\bf p}^2/2m + V_{eff}({\bf r}) - \mu ] / k_B T} -1 \right)^{-1}
\label{SCHF2}  \\
&&n_{th}({\bf r}) = \frac{1}{\lambda_T^3} \;\; 
g_{3/2}\left( e^{\left[\mu - V_{eff}({\bf r}) \right] /k_B T } \right )
\label{SCHF3}  \\
&&V_{eff}({\bf r}) = V_{tr}({\bf r}) + 2\,  g\,  n_c({\bf r}) + 2\,  g\,  n_{th}({\bf r})
\\
&&\mu = g\, n_c(0) + 2\, g\, n_{th}(0)  \,,
\end{eqnarray}
where
\begin{equation}
\lambda_T = \left( \frac{2\pi\hbar^2}{mk_B T} \right)^{1/2} 
\end{equation}
is the thermal de Broglie wavelength and the $g_{3/2}(z)$ function is given by the expansion:
\begin{equation}
g_{3/2}(z) = \sum_{n=1}^\infty \frac{z^n}{n^{3/2}}  \,.
\end{equation}

The basic variables in this approach are the condensate density, $n_c({\bf r})$, and the
distribution function in phase space for thermal atoms, $f({\bf r},{\bf p})$. They are
calculated according to the Eqs. (\ref{SCHF1}) and (\ref{SCHF2}). The thermal density, 
$n_{th}({\bf r})$, is just an integral of the distribution function $f({\bf r},{\bf p})$ over 
momenta and can be found analytically (Eq. (\ref{SCHF3})). Since both the effective potential, 
$V_{eff}({\bf r})$, and the chemical potential, $\mu$, appearing on the right-hand sides of 
Eqs. (\ref{SCHF1}), (\ref{SCHF2}), and (\ref{SCHF3}), depend on the condensate and thermal 
densities the set of Eqs. (\ref{SCHF1}) and (\ref{SCHF2}) is well suited to be solved iteratively. 
For that, however, we have to first choose the temperature of the
system and then keep fixed the total number of atoms (which is 
$N=\int d{\bf r}\, (n_c({\bf r}) + n_{th}({\bf r}))$) during the iterations.
Having the densities of condensed and noncondensed fractions one can
easily calculate two important parameters: the condensate fraction and the total energy 
per particle. Now the strategy is as follows: find the input parameters $(N,T)$ in
such a way that the condensate fraction and the total energy per atom calculated
within the Hartree-Fock method match the values calculated from the classical field at equilibrium. This procedure allows one to determine
the number of atoms and the temperature assigned to the classical field separately.

\begin{figure}[htb]
\resizebox{3.2in}{2.1in} {\includegraphics{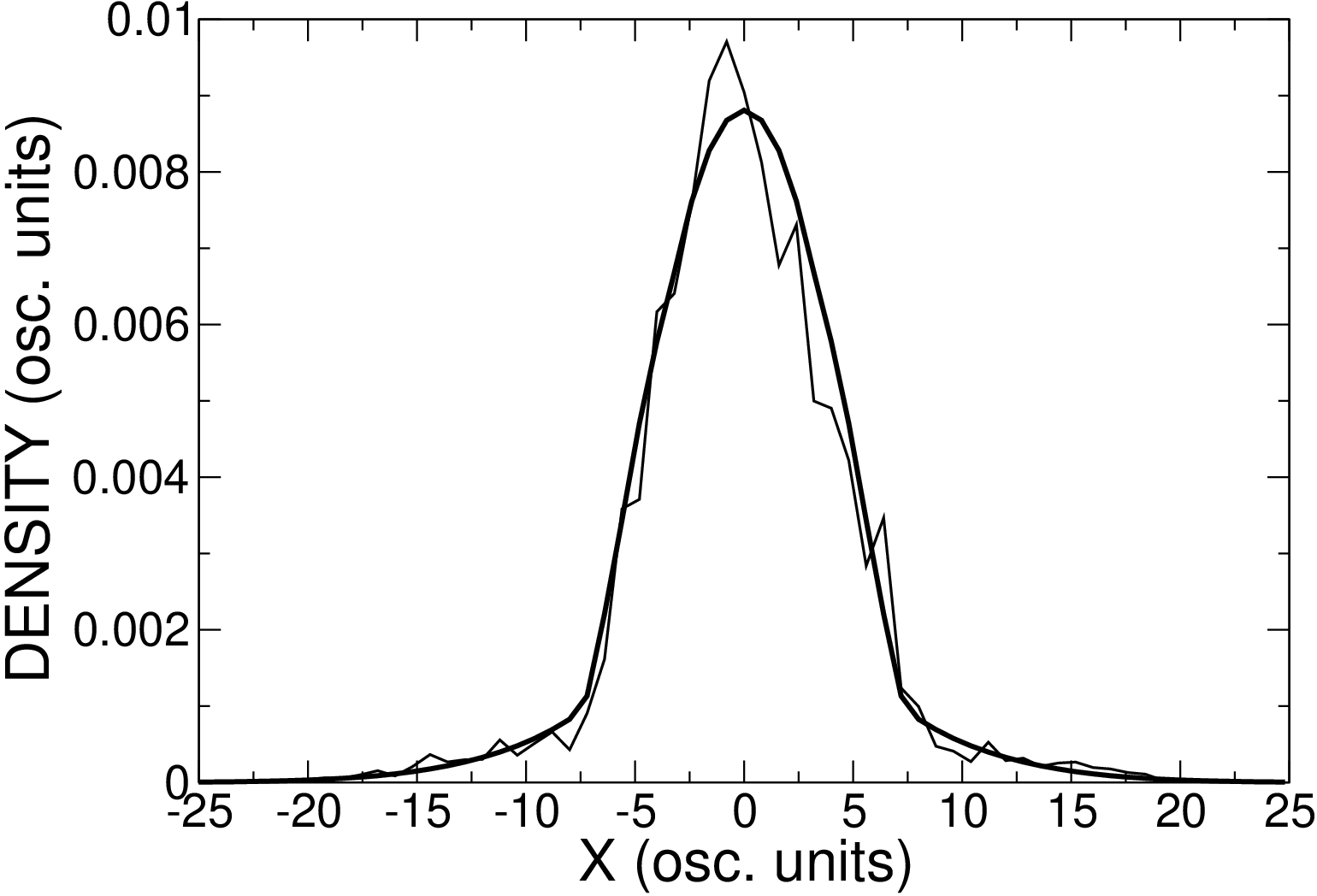}}
\caption{Total density cuts for the self-consistent Hartree-Fock (thick line) and
classical field (thin line) methods for a system with the total number of atoms
$N=17306$ and the temperature $T=128.7\,$nK.}
\label{SCHF-CFA}
\end{figure}

These parameters for the example discussed in the context of Figs. \ref{relpop}
and \ref{kTN} are found to be $N=17306$ and $T=128.7\,$nK. These values are very close to what was obtained  in the previous section with the cutoff 
occupation $N_{cut}=0.46$, and differ by $3.7\%$ in the temperature and $12.8\%$ in the total number of atoms.
The agreement is good even though the average occupation of the
highest energy modes (the last modes considered in the classical field approximation
as being macroscopically occupied) was taken the same as for an ideal 
gas. Unfortunately, there is no data available for the average occupation of the cutoff region modes 
for the weakly interacting gas considered here. Changing slightly this
cutoff occupation number, the agreement between both approaches to the system parameters can be made
even better (for example, for $N_{cut}=0.48$ the difference is $1.3\%$ and $7.4\%$ in
the temperature and the total number of atoms, respectively). 
In Fig. \ref{SCHF-CFA} the total Hartree-Fock and classical field densities
are plotted together for parameters as in Figs. \ref{relpop} and \ref{kTN} showing
a good agreement. The classical field density, however, exhibits all the fluctuations which are not
present in the Hartree-Fock model.

In Table \ref{tab1} we again compare the parameters of the system (total
number of atoms and temperature) for the methods described in this and in the previous
sections. However, this time the comparison is for several different equilibrium states. All the states being compared 
are used later in the simulations related to JILA experiment \cite{Cornell-osc} (see Secs. 
\ref{mode0} and \ref{mode2}).  
\begin{table}[htb]
 \begin{tabular}{|c||l|l|}
\hline
  $ $ & $CFA$ & $HF$  \\
\hline \hline
$T$[nK] &  79.8    &    91.8     \\
$N$     &  6751    &    11300    \\
\hline
$T$[nK] &  124.1   &    128.7    \\
$N$     &  15342   &    17306    \\
\hline              
$T$[nK] &  142.1   &    153.9    \\
$N$     &  16347   &    18699    \\
\hline                
$T$[nK] &  154.4   &    168.8    \\
$N$     &  19313   &    21509    \\
\hline      
$T$[nK] &  177.8   &    190.4    \\
$N$     &  26984   &    27875    \\
\hline      
$T$[nK] &  195.3   &    204.8    \\
$N$     &  27770   &    31222    \\
\hline     
$T$[nK] &  264.1   &    240.9    \\
$N$     &  61876   &    43572    \\
\hline     
 \end{tabular} 
\caption{Comparison (for several final equilibrium states as per Table \ref{tab2}) between the results (the temperature and the total number of atoms)
obtained by projecting the classical field on the harmonic oscillator states as described
in Sec. \ref{eqstates} ($CFA$ column) and by utilizing the self-consistent Hartree-Fock
method according to Sec. \ref{HFmethod} ($HF$ column). For $CFA$ data $N_{cut}=0.46$. }
\label{tab1}
\end{table}

\subsection{Obtaining equilibrium states on demand}
\label{ondemand}

In Secs. \ref{eqstates} and \ref{HFmethod} we detailed the methods for retrieving
the total number of atoms and the temperature assigned to the classical field
at equilibrium. Since the product $gN$ is an initial parameter it means that the 
coupling constant $g$ (and consequently the scattering length $a$) is known only after the 
classical field is thermalized. Unfortunately, it usually differs from the value that was 
used to calculate an initial value of $gN$. 
Therefore, an approach for obtaining the classical field at equilibrium corresponding to 
given values of the total number of atoms and the temperature is required.

The method we have developed is rather simple although demanding from a numerical point
of view. Let's assume that we need the classical field  which at equilibrium describes
the system with given parameters $N$ and $T$. We start from a solution obtained
from the self-consistent Hartree-Fock method corresponding to the chosen parameters.
Then we build an initial classical field as follows:
\begin{equation}
\Psi ({\rm {\bf r}},t=0) = \sqrt{n_c({\bf r}) } + \sqrt{n_{th}({\bf r})}\, e^{i \varphi({\bf r}) } 
\label{psi}
\end{equation}
and randomize the phase $\varphi({\bf r})$ and the density $n_{th}({\bf r})$ in such a way that the 
total energy per atom
in the classical field equals the corresponding energy in the Hartree-Fock model.
The presence of the phase factor in the second term in (\ref{psi}) is necessary.
Without this, the classical field suffers from a lack of kinetic energy in comparison with
the Hartree-Fock method, where it is calculated from
the distribution function $f({\bf r},{\bf p})$:
\begin{eqnarray}
&&E_{kin} = \frac{1}{h^3} \int d{\bf r}\, d{\bf p}\, \frac{{\bf p}^2}{2m}
f({\bf r},{\bf p})    \nonumber  \\
&&= \frac{3 k_B T}{2 \lambda_T^3}
\int d{\bf r}\,   g_{5/2}\left( e^{\left[\mu - V_{eff}({\bf r}) \right] /kT } \right )          
\,.
\label{kinene}
\end{eqnarray}

Now we evolve the classical field according to Eq. (\ref{CFequation}) and let the field 
thermalize. During the thermalisation, the total energy per atom is a constant of
motion, but the condensate fraction usually changes. However, there will be a particular value of the 
spatial step of the grid for which the condensate fraction does not change in time. 
Then, since the total energy per atom is a constant of motion the values of 
parameters $n_0$ and $E_{tot}/N$ at the end of thermalization process are the same as at the 
beginning. Consequently, the number of particles $N$ and the
temperature $T$ must be the same as chosen at the beginning as well. Although, the procedure just 
described is numerically time consuming (since it requires several trials to obtain the proper
value of the spatial step), 
 it is much more efficient than attempts to match final $T$ and $N$ simultaneously, which require 
search in two-dimensional parameter space.
Here, the energy matching is computationally fast because it requires no thermalisation, while the final 
$T,N$ matching is done with only one free parameter, the spatial lattice spacing.

\section{Results for the m=0 mode}
\label{mode0}

Our numerical procedure involves the following steps: First, we find the classical field
(as described in \ref{ondemand}) corresponding to the Bose gas at equilibrium confined in a 
harmonic trap with frequencies $\omega_{\bot} \equiv \omega_{x,y} = 2\pi \times 129\,$Hz and 
$\omega_z = 2\pi \times 365\,$Hz. According to the experiment \cite{Cornell-osc} the total
number of atoms was of the order of ten to a few tens of thousands and the initial temperature was
ranging up to the critical temperature. The number of condensed atoms remained on a level of
several thousand. The numerical values of $N$, $N_0$, $T$, and the reduced temperature
$T^{\prime} \equiv T/T_c$ (with $T_c$ being the transition temperature for a harmonically
confined ideal gas) for the states used in the simulations are shown in Table \ref{tab2}. 

\begin{table}[htb]
 \begin{tabular}{|c|c|c|c|}
\hline
  $N$ & $N_0$ & $T$[nK] & $T^{\prime}$ \\
\hline \hline
11300 & 8558 & 92.8 & 0.497   \\
17306 & 10349 & 128.7 & 0.605 \\
18700 & 7905 & 153.9 & 0.705  \\
21509 & 7768 & 168.8 & 0.738  \\
27875 & 8496 & 190.4 & 0.763  \\
31222 & 7753 & 204.8 & 0.791  \\
43572 & 7111 & 240.9 & 0.832  \\
\hline
 \end{tabular} 
\caption{ Numerical values of $N$, $N_0$, $T$, and $T^{\prime}$ used in the simulations.}
\label{tab2}
\end{table} 

Next, the Bose gas is disturbed by a sinusoidal perturbation to the trapping potential. Since 
the classical field describes both the condensed and noncondensed atoms, the disturbance
of the classical field means that both components are simultaneously perturbed. To excite 
the $m=0$ and the $m=2$ quadrupole modes, the perturbation of the trapping potential takes 
the form:
\begin{eqnarray}
\delta V_{tr}({\rm {\bf r}},t) = A(t)\, [\omega_x^2 x^2 \sin(\omega_d\, t + \phi) + 
\omega_y^2 y^2 \sin(\omega_d\, t)]  \,,   \nonumber  \\ 
\label{pertur}
\end{eqnarray}
where $\omega_d$ is the driving frequency and $\phi$ is a phase shift between the
$x$ and $y$ direction perturbations. The choice of the phase shift $\phi$ determines
the symmetry of the excited collective mode - for $\phi =0$ ($\phi =\pi$) the $m=0$ ($m=2$) 
mode is excited. The perturbation, as in the experiment, lasts for $14\,$ms and the amplitude 
$A(t)\, (=0.05)$ takes small value to avoid any nonlinear effects.

\begin{figure}[htb]
\resizebox{2.9in}{3.2in} {\includegraphics{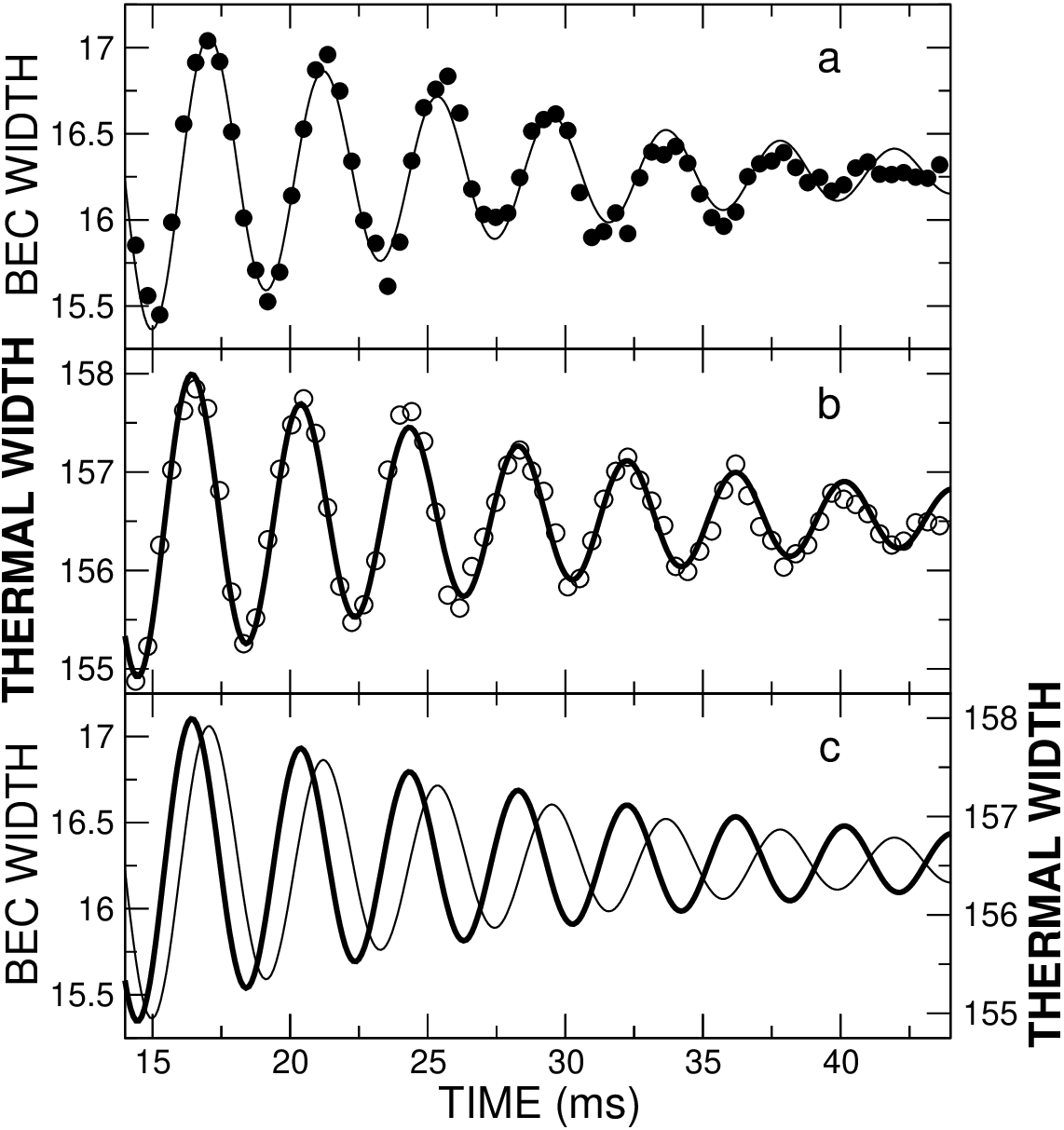}}
\caption{Widths of a condensate (upper frame) and its thermal cloud (middle frame) as a function of 
time for the $m=0$ mode. The widths (solid and open symbols) are calculated from the formula (\ref{width}) 
after the condensate and the thermal cloud are extracted out of the classical field. The solid lines 
are fits to an exponentially damped sine function:
$A_{c,th}\; exp(-\gamma_{c,th} t)\, sin(\omega_{c,th} t + \vartheta_{c,th}) + B_{c,th}$. 
These fits allow us to obtain the frequencies and the damping rates of the oscillations of the condensate and 
the thermal cloud in response to the external perturbation. 
The initial temperature of the cloud and the driving frequency are $T'=0.8$ and 
$\omega_d=1.75\, \omega_{\bot}$.  
The lower frame shows more clearly the frequency and phase shifts between the condensate and the thermal cloud.}
\label{oscillations}
\end{figure}

After the perturbation is turned off, the classical field is oscillating in time. Is it possible
that the condensed and noncondensed components (extracted from the single classical
field) exhibit oscillations with different frequencies? To answer these question  we split the
classical field into the condensate and the thermal cloud in the way described in Sec. \ref{formalism} 
and calculate the widths of both components from the formula:
\begin{eqnarray}
w_{c,th} = \int dx\, dy\, (x^2 + y^2)\, n_{c,th}(x,y)       \,.
\label{width}
\end{eqnarray}
Results are shown in Fig. \ref{oscillations} for the $m=0$ mode. Solid symbols (upper frame) represent
the condensate widths, whereas the open symbols (middle frame) stand for thermal cloud widths.
Solid lines in both frames are fits by an exponentially damped sine waves:
\begin{eqnarray}
A_{c,th}\; \exp(-\gamma_{c,th} t)\, \sin(\omega_{c,th} t + \vartheta_{c,th}) + B_{c,th} 
\label{fit}
\end{eqnarray}
to numerical data. As in the experiment, fits are performed based on five initial oscillations.
The lower frame shows, indeed, that condensate and thermal components oscillate with
different frequencies. Moreover, these oscillations are phase shifted  and the condensate oscillates slower than the thermal cloud.

\begin{figure}[thb]
\resizebox{3.2in}{2.4in} {\includegraphics{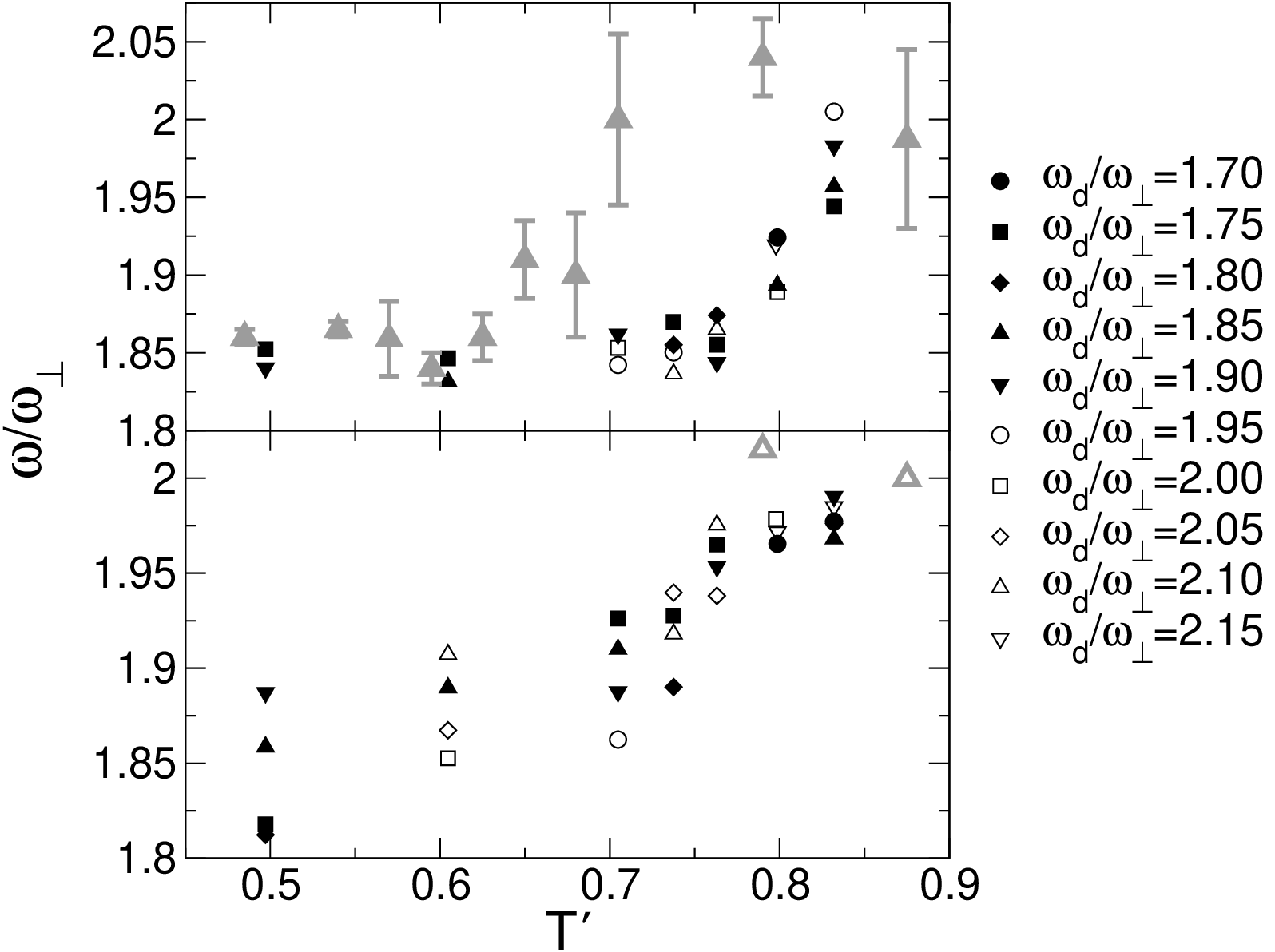}}
\caption{Condensate (upper frame) and the thermal cloud (lower frame) frequencies for the $m=0$ 
collective mode as a function of reduced temperature. Black (solid and open) symbols correspond 
to the numerical results obtained for various driving frequencies (according to the legend) 
whereas gray symbols (solid for the condensate and open for the thermal cloud) with error bars 
are taken from the experiment by Jin {\it et al.} \cite{Cornell-osc}.  }
\label{m0fr}
\end{figure}

\begin{figure}[thb]
\resizebox{3.2in}{2.4in} {\includegraphics{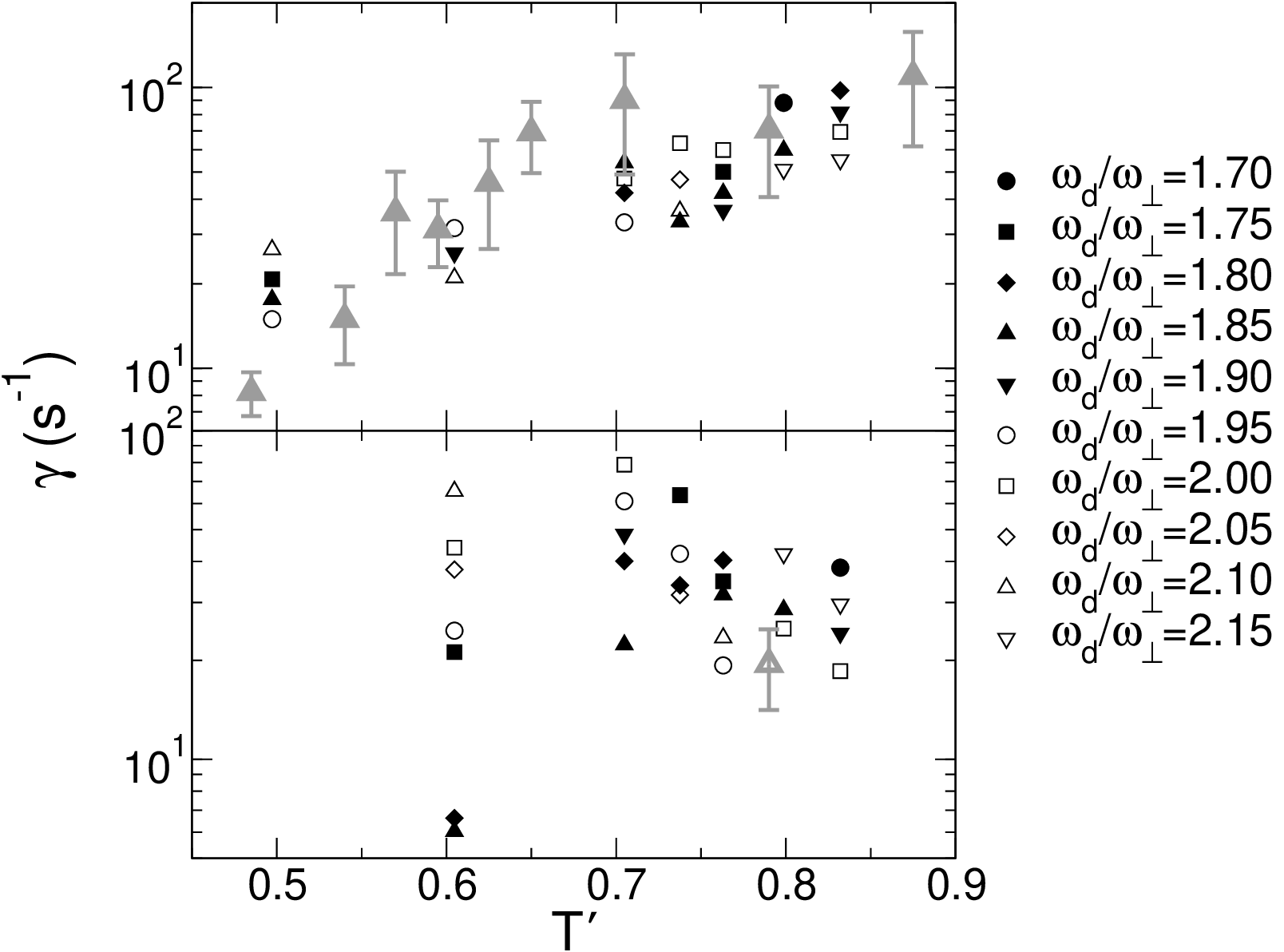}}
\caption{Damping rates for the condensate (upper frame) and the thermal cloud (lower frame) 
for the $m=0$ mode as a function of reduced temperature. Black (solid and open) symbols correspond 
to the numerical results obtained for various driving frequencies (according to the legend)
whereas gray symbols (solid for the condensate and open for the thermal cloud) with error 
bars are taken from the experiment by Jin {\it et al.} \cite{Cornell-osc}.}
\label{m0dr}
\end{figure}

In Figs. \ref{m0fr} and \ref{m0dr} we summarize our results for the $m=0$ mode. In Fig. \ref{m0fr}
we plot the frequencies of  the condensate and thermal fractions response to the external perturbation
for various temperatures together with the experimental data of Ref. \cite{Cornell-osc}. Black solid 
and open symbols represent data for a condensate (upper frame) and a thermal cloud (lower frame)
for various driving frequencies according to the legend attached to the figure. Gray symbols with 
error bars are the experimental results. Up to temperature $T \approx 0.6 T_c$ both components oscillate
with the same frequency, which is the natural condensate 
frequency for the $m=0$ collective mode. At approximately $0.65 T_c$ a rather sudden upward shift in 
condensate frequency is observed in the experiment. Our numerics shows that at this temperature the
dynamics of the thermal cloud changes. The thermal component starts to oscillate with a higher
frequency approaching eventually $2 \omega_{\bot}$ which is the oscillation frequency of a thermal 
gas alone. For higher temperatures the thermal fraction becomes dominant and finally thermal atoms
change the dynamics of the condensate in such a way that condensed atoms oscillate along with
the thermal ones. Unfortunately, the influence of the thermal cloud on the condensate is apparently 
too weak and consequently the condensate starts to oscillate with higher frequencies only for
higher temperatures.

Our results also show that the notion of natural condensate frequency breaks down when the thermal
cloud is present. The condensate response depends on the dynamics of the thermal cloud and, in fact,
the possible frequencies for the condensate oscillation lie in an interval which gets wider for
higher temperatures. In particular, no two branches of frequencies are visible as reported in 
Ref. \cite{Zaremba-osc}. In Fig. \ref{m0dr} we compare the numerical and experimental damping rates 
for the oscillations of the condensed and noncondensed components. There is some discrepancy for
higher temperatures where the thermal component seems to be damped too strongly in comparison with
the experimental data. Perhaps this increases the temperature at which the frequency of the 
condensate fraction shifts up.

\section{Results for the m=2 mode}
\label{mode2}
We now do the same analysis for the $m=2$ mode. The system is disturbed by changing the trapping
potential according to (\ref{pertur}) with the phase shift $\phi =\pi$. Such a perturbation
excites the quadrupole $m=2$ oscillations. The classical field is split
into the condensed and thermal parts and the condensate 
and thermal widths are calculated using formulas exhibiting the $m=2$ mode's symmetry:
\begin{eqnarray}
w_{c,th} = \int dx\, dy\, (x^2 - y^2)\, n_{c,th}(x,y)   \,.   
\label{width2}
\end{eqnarray}
As before, these data are fitted by exponentially damped sine waves.

\begin{figure}[thb]
\resizebox{3.2in}{2.1in} {\includegraphics{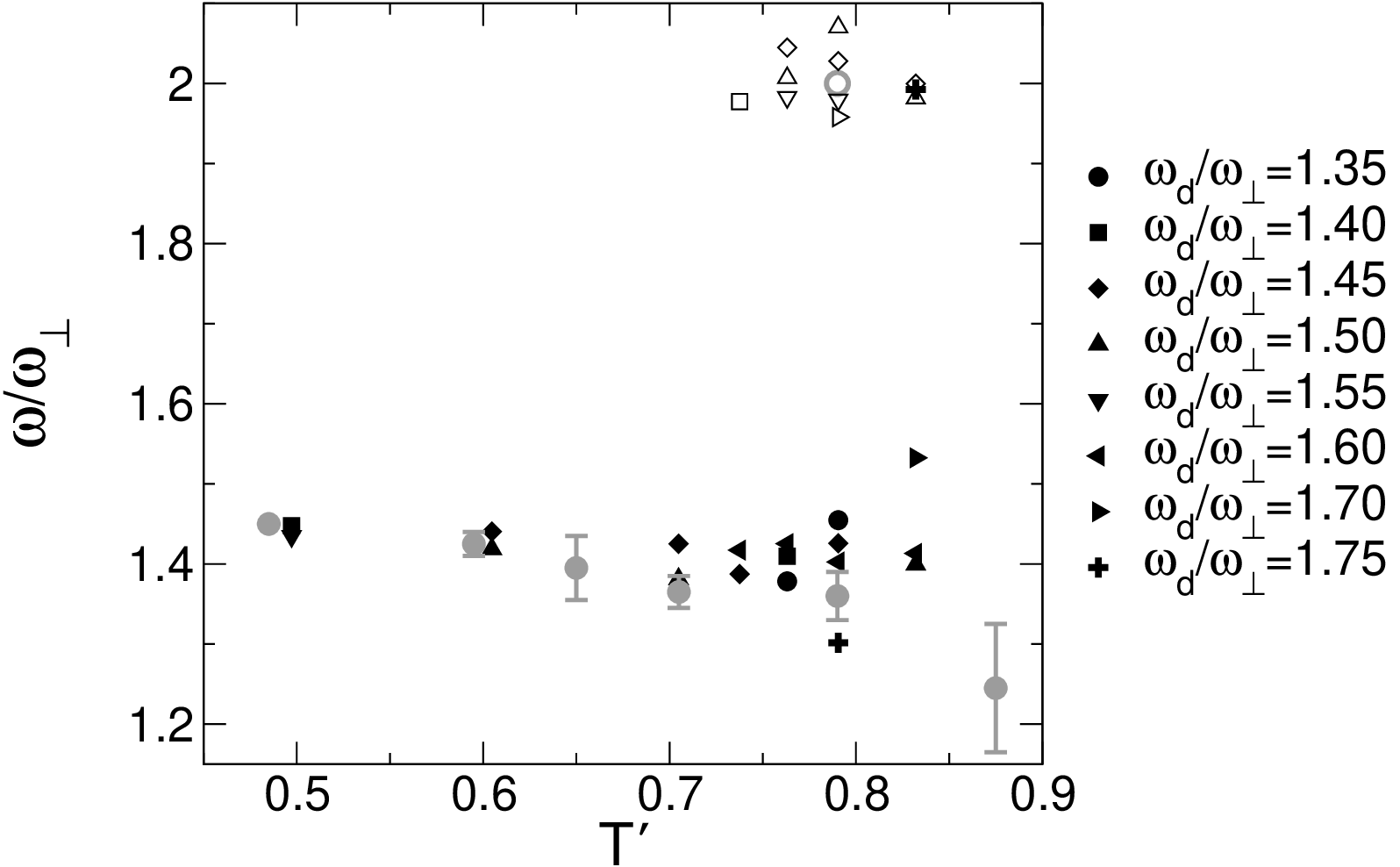}}
\caption{Condensate (solid symbols) and thermal cloud (open symbols) frequencies 
for the $m=2$ collective mode, as a function of reduced temperature. Black symbols correspond to
the numerical results obtained for various driving frequencies (according to the legend,
here the shape of the symbol determines the driving frequency), whereas gray symbols with error 
bars are taken from the experiment by Jin {\it et al.} \cite{Cornell-osc}. 
Note, however, that the only experimental point for the thermal cloud is marked with an
open gray circle.}
\label{m2fr}
\end{figure}

\begin{figure}[thb]
\resizebox{3.2in}{2.4in} {\includegraphics{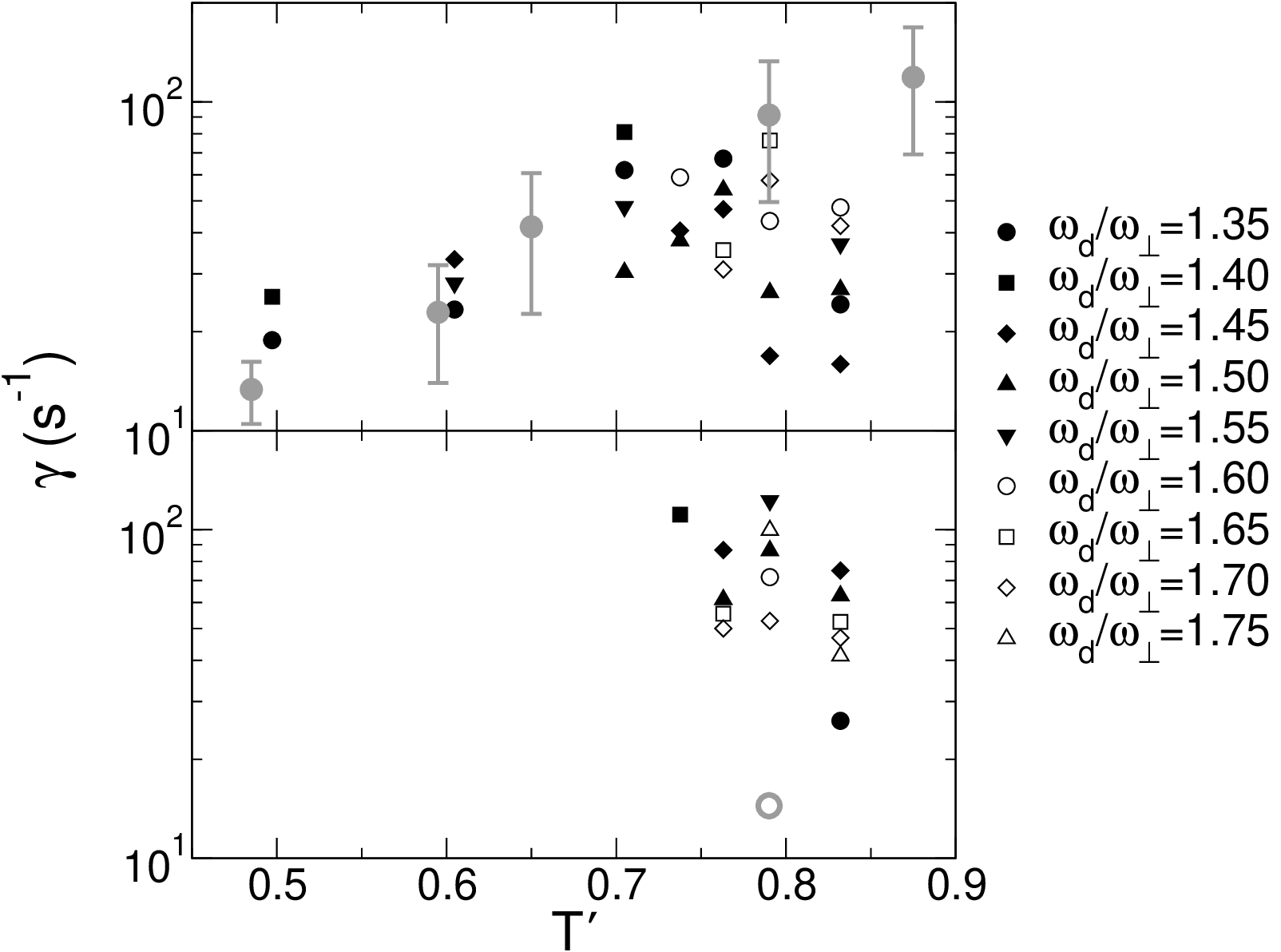}}
\caption{Damping rates for the condensate (upper frame) and the thermal cloud (lower frame) 
for the$m=2$ mode, as a function of reduced temperature. Black (solid and open) symbols correspond 
to the numerical results obtained for various driving frequencies (according to the legend),
whereas gray symbols (solid for the condensate and open for the thermal cloud) with error 
bars are taken from the experiment by Jin {\it et al.} \cite{Cornell-osc}. Note that 
the only experimental point for the thermal cloud is marked with an open gray circle. }
\label{m2dr}
\end{figure}

Frequencies and damping rates for the condensate and the thermal cloud are plotted in Figs. 
\ref{m2fr} and \ref{m2dr}. Again, the system responds in a way that depends on the driving 
frequency. A comment on how the authors
of the experiment \cite{Cornell-osc} choose the driving frequency is in place here. They say that the
driving frequency is set to match the frequency of the excitation being studied. This could make 
sense for $m=0$ mode where supposedly the condensed and noncondensed fractions always oscillate
in phase. Supposedly, because in \cite{Cornell-osc} we see only two data points (both having frequency approximately
$2 \omega_{\bot}$) showing the behavior of the thermal cloud in the $m=0$ mode. However, for
the $m=2$ mode the frequencies at which the condensate and thermal cloud oscillate are different,
and the meaning of the driving frequency as the frequency of the excitation being studied is 
not clear. Therefore, in Fig. \ref{m2fr} we show our data for various driving frequencies.
Good agreement with experiment is found for appropriate driving frequencies up to temperatures
$\approx 0.8 T_c$. For temperatures approaching the critical one, however, we observe the same
effect as for the $m=0$ mode. The thermal cloud becomes dominant and the frequencies at which the condensate
oscillates become higher. Fig. \ref{m2fr} shows that already for a temperature $T = 0.83 T_c$, with 
the driving frequency $\omega_d = 1.75 \omega_{\bot}$, the condensate oscillates with the
frequency of the thermal cloud. Such a behavior was not observed experimentally.
On the other hand, it is an expected behavior, since the condensed fraction becomes smaller and smaller
while the critical temperature is approached, and the dynamics should be dominated by the
thermal cloud.
In the case of the thermal cloud, our model predicts frequencies in agreement with the
experiment (other theoretical studies of the JILA experiment either do not show results for the
thermal atoms for the $m=2$ mode, or predict frequencies that are very different than observed
experimentally).

In Fig. \ref{m2dr} we compare the numerical and experimental damping rates for the oscillations 
of the condensed and noncondensed parts of the system. As for the $m=0$ mode, there is some discrepancy 
for higher temperatures where the thermal cloud is damped too strongly, whereas the condensate 
is damped too weakly.

\section{Conclusions}
\label{concl}
In conclusion, we have presented in detail the construction of the classical field 
describing the desired number of atoms, confined in any trapping potential, at a prescribed
temperature. We have studied the oscillations of the Bose-Einstein condensate in the 
presence of a thermal cloud. As in the experiment, we find the temperature dependent condensate 
frequency shift for both the $m=0$ and $m=2$ collective oscillation modes. For the $m=0$ mode, 
the thermal atoms pull the condensate fraction along, and above some characteristic 
temperature $(\approx 0.8 T_c)$ the condensate tends to oscillate with the frequency of the 
thermal part (approximately equal to $2 \omega_{\bot}$). Unfortunately, in the present 
version of the classical field approximation, the value of this characteristic temperature turns 
out to be about $20 \%$ higher than the one observed in the experiment.
For the $m=2$ mode, on the other hand, the frequency at which the condensate oscillates first decreases with 
an increase of temperature (the thermal cloud oscillating with its natural frequency equal to 
$2 \omega_{\bot}$ damps the condensate motion) but when the temperature gets closer to the 
critical temperature the condensate starts to oscillate with higher frequencies, approaching
the frequency of the thermal cloud.

\acknowledgments
We are grateful to Piotr Deuar for his critical reading of the manuscript and his
valuable comments.
The authors acknowledge support by Polish Government research funds for 2009-2011.
Some of the results have been obtained using computers at the Interdisciplinary
Centre for Mathematical and Computational Modeling of Warsaw University. 
The Centre for Quantum Technologies is a Research Centre of Excellence funded by the Ministry of Education
and the National Research Foundation of Singapore.

\end{document}